\documentclass{aastex}
\usepackage{emulateapj5}
\shorttitle{Inclusion of turbulence in solar modeling}
\newcommand{\p}[2]{\frac{\partial #1}{\partial #2}}

\newcommand{\od}[2]{\frac{d #1}{d #2}}
\newcommand{\s}[1]{\mbox{\scriptsize{#1}}}
\begin{document}

\title{Inclusion of turbulence in solar modeling}

\author{L. H. Li, F. J. Robinson, P. Demarque, S. Sofia}
\affil{Department of Astronomy, Yale University, P.O. Box 208101, New Haven,CT 06520-8101}
\and
\author{D. B. Guenther}
\affil{Department of Astronomy and Physics, Saint Mary's University, 
Halifax,
Nova Scotia B3H 3C3, Canada}

\begin{abstract}
The general consensus is that in order to reproduce the observed solar p-mode oscillation
frequencies, turbulence should be included in solar models.
However,  until now there has not been any well-tested efficient method to incorporate
turbulence into  solar modeling. We present here two methods to include turbulence 
in solar modeling within the framework of the mixing length theory, using the 
turbulent velocity obtained from numerical simulations of the highly superadiabatic 
layer of the sun at three stages of its evolution. The first approach  is to include the
turbulent pressure alone, and the second is to include both the turbulent
pressure and the turbulent kinetic energy. The latter is achieved by introducing two
variables: the turbulent kinetic energy per unit mass, and the effective 
ratio of specific heats due to the turbulent perturbation. These are treated as 
additions to the standard thermodynamic coordinates (e.g. pressure and temperature). 
We investigate the effects of both treatments of turbulence on the structure 
variables, the adiabatic sound speed, the structure of the highly superadiabatic layer, and 
the p-mode frequencies. We find that the second method reproduces the SAL structure 
obtained in 3D simulations, and produces a p-mode frequency correction 
an order of magnitude better than the first method.
\end{abstract}

\section{Introduction}\label{s1}

The effects of turbulence on the structure of a solar model depends not only on
how it is modeled, but also on how it is incorporated into the
solar model. The change in the p-mode oscillation frequencies caused by turbulence can
be used as a measure of the effects of turbulence on the structure of solar
models, since helioseismology provides an opportunity to probe directly and
sensitively solar structure. In the standard solar models (SSM), the turbulent convection is
modeled in terms of the local mixing-length theory (MLT), but the turbulent pressure,
turbulent kinetic energy and turbulent entropy are ignored. The fact that the
computed frequencies of p-modes from standard solar models are higher than the
observed values, shows that standard solar models need to be refined. The
frequency dependence of the discrepancy reveals that there is a problem with the
model, and that the problem lies very near the surface. Balmforth (1992a, 1992b,
1992c) uses the non-local mixing-length theory to model the turbulent
convection, and includes the turbulent pressure, but ignores the turbulent kinetic
energy and turbulent entropy in modeling the sun. The computed frequencies of
p-modes from such a model are even higher than those computed from the standard
solar models (Balmforth 1992a). Canuto (1990,1996) has proposed a semi-analytical
model of the turbulent convection. The main idea is to include a full turbulent
spectrum in the model of convection. Applications to stellar models have been
discussed by Canuto \&\ Mazzitelli (1991,1992) and Canuto, Goldman, \&\
Mazzitelli (1996). The parameters of the Canuto-Mazzitelli (C\&M) approach are
derived from the laboratory experiments of the incompressible convection, and
extrapolated to the stellar conditions. Using this approach, the superadiabatic 
peak is much higher than that of the standard solar models, while the computed
frequencies of solar p-modes are closer to the observed values than those from
the standard solar models.

Turbulence is a highly nonlinear phenomenon. Numerical experiments
are a direct, effective way to investigate turbulence, in addition to the laboratory
experiments \cite{NSSD00}. Chan \&\ Sofia (1989) performed a three-dimensional numerical
simulation of the deep compressible convection, where the superadiabatic excess in
the temperature gradient is very small. They showed that in the deep regions, the
mixing-length approximation is valid. They derived expressions for key
physical parameters such as the convective energy flux. These, in principle, can
be conveniently applied to stellar models. The Chan-Sofia formula for the
convective flux was incorporated into the Yale stellar evolution code by Lydon
(1993) and Lydon, Fox, \&\ Sofia (1992). The diffusion approximation was used to
treat radiative transfer, and the radiative atmosphere was treated by Lydon et
al. (1992) as in the standard MLT Yale models (e.g., Guenther et al. 1992). They
found that the peak of the superadiabatic layer (SAL) is not as high as in the C\&M
model and is located at a greater depth below the surface. However, their models
yielded a larger discrepancy between the observed solar p-mode frequencies and those 
computed from standard solar models. Part of the discrepancy could be attributed to the fact
that these models did not match the solar radius very precisely. But this is expected 
since the simulation, which is valid for the deeper adiabatic regions, is 
extrapolated into the SAL, where the temperature gradient greatly exceeds 
the adiabatic gradient, and some inaccuracy is inevitable.

In order to overcome these difficulties, Kim (1993) and Kim et al. (1995, 1996)
conducted a three-dimensional numerical simulation whose domain includes 
shallower layers. This simulation treats the coupling of radiation and
convection and includes realistic equation of state and
radiative opacities, taken from the Yale Stellar Evolution Code. The Kim et al.'s
(1995, 1996) simulation treats radiative transfer in the diffusion
approximation, which is only valid in an optically thick medium, and
consequently cannot be used in the solar atmosphere. Thus the top boundary of
the simulation was set below the SAL peak. The Kim et al.'s (1996) models were
parameterized as a varying mixing length with depth by Demarque, Guenther, \& Kim
(1997), in precisely calibrated solar models. The p-mode frequencies of the
models were found to agree more closely with the observed solar p-mode
frequencies, than the standard solar model frequencies, but these models
exhibit, like the C\&M models, a higher SAL peak than the standard solar
models.

More recently, Kim \&\ Chan (1998) have completed a three-dimensional radiative
hydrodynamic simulation of the complete extent of the SAL including the solar
atmosphere (about 2 pressure scale heights above and 2.5 pressure scale heights
below the SAL). The numerical approach was described by Kim \&\ Chan (1997). The
simulation of Kim \&\ Chan is fully compressible and uses the realistic equation of 
state and opacities. The radiation has been treated by utilizing the three-dimensional
Eddington approximation, which is valid in both the optically thin regions near the
surface and the optically thick regions in the interior. Demarque, Guenther, \&\ 
Kim (1999) mimicked this simulation in the calibrated solar models by increasing 
the opacity coefficient $\kappa$ near the surface. Such models are called perturbed-$\kappa$ models. 
The discrepancy between calculated and observed p-mode frequencies decreases 
when compared to standard solar models. A deeper layer  (about 5 pressure 
scale heights above and 6 pressure scale heights below the SAL) was simulated 
by Stein \&\ Nordlund (1998) using different numerical methods for the convective 
and radiative components. Instead of parameterizing the simulation, Rosenthal et al. (1999)
matched the simulation to an envelope which was constructed using an MLT
envelope code. The computed frequencies are in better agreement with the
observed solar p-mode frequencies when compared to the standard solar models. Abbett
et al. (1997) have also discussed the same transition layer.

One way to include turbulence in stellar models is to use the numerical 
simulations to directly calculate the convective temperature gradient and its 
derivatives needed in stellar model calculations. However, because turbulence 
is chaotic, nonlocal, and three-dimensional, and because it involves nonlinear 
interactions over many disparate length scales, the simulation for the whole 
convective zone is too computationally expensive for stellar model calculations.
Besides, the accuracy of the simulation, if performed, is too
low to match the required accuracy for the stellar model calculations. Fortunately,
the existing simulations show that the MLT prediction deviates from 
simulations only in the SAL part of the convection zone. Even so, it is still
impractical to calculate the convective temperature gradient and its derivatives,
from numerical simulation of the SAL in the stellar evolution model caculations.
Robinson et al. (2000) have performed some hydrodynamical simulatons 
(the viscosity parameter $c_\mu=0.2\sqrt{2}$, see below) of the SAL at three stages in 
the solar evolution: the zero-age main sequence (ZAMS), the present 
sun and the subgiant. The results show that the turbulent velocities in the SAL, as
functions of gas pressure, change little for all these stages. Therefore, it is 
feasible to compute the effects of turbulence on the 
convective gradients. Overshooting was observed in both the solar subatmosphere 
and 3D numerical simulations, and the $\kappa$-models \cite{DGK99} demonstrate that
overshooting is likely to be one of the keys to match the real Sun. It is a 
challenge to include overshooting within the framework of the local MLT since
overshooting is a nonlocal phenomenon.

In this paper, we use the results of the 3D numerical simulations of 
turbulence to calculate the convective temperature gradient, and its derivatives 
needed in solar model calculations. As MLT is valid in most of the convective zone, 
it is convenient to include turbulence within the framework of MLT. In \S\ref{s2}
we summarize the new results ($c_\mu=0.2$) of the 3D simulations of 
the solar SAL, at 3 stages of its evolution. These simulations are similar to 
those by Robinson et al. (2000), but with lower viscosities. \S\ref{s3} describes how
to calculate the convective temperature gradient using turbulent velocities. We
describe the calibrated solar models with turbulence in \S\ref{s4}. In
\S\ref{s5} the influence of turbulence on the structure variables, the adiabatic 
sound speed, the structure of the highly superadiabatic layer, and 
the p-mode frequencies are calculated and compared to the observed solar 
p-mode frequencies. Concluding remarks are presented in the last section.

\section{Turbulent velocities}\label{s2}

We incorporate the radiative hydrodynamical simulations 
of the outer layers of the sun into the 1D stellar models. Three 3D simulations 
have been performed using the hydrostatic 1D stellar models, at three stages of 
its evolution (ZAMS, present sun and subgiant), as starting points. The physics 
(thermodynamics, the equation of state, and opacities) in the simulation is 
the same as in the 1D stellar models. These simulations follow closely the approach 
described by Kim \& Chan (1998), and are described in more detail by 
Robinson et al. (2000). The full hydrodynamical equations were solved in a thin 
subsection of the stellar model, i.e. a 3D box located in the vicinity of the 
photosphere. For the radiative transport, the diffusion approximation was used in the deep 
region ($\tau > 10^3$) of the simulation, while the 3D Eddington approximation 
was used (Unno \& Spiegel 1966) in the region above. After the simulation 
had reached a steady state, statistical integrations were performed for each
simulation for over 2500 seconds in the case of the solar surface
convection.

Turbulence can be measured by the turbulent Mach number ${\cal M}=v''/v_s$,
where $v''$ is the turbulent velocity, and $v_s$ is the sound speed. The MLT is valid
when ${\cal M}$ is sufficiently small. In the outer layers of the sun ${\cal M}$
can be of order unity (Cox and Giuli 1968), but in the deep convection region
${\cal M}$ is almost zero. The turbulent velocity is defined by the velocity variance:
\begin{equation}
  v''_i = (\overline{{v_i}^2}-\overline{{v_i}}^2)^{1/2}, \label{eq:rms}
\end{equation}
where the overbar denotes a combined horizontal and temporal average, and $v_i$ is the total 
velocity. Fig.~\ref{mach} shows the run of ${\cal M}$ as a function of $\log P$ (in base 10)
in the convection simulations for the ZAMS model, the present sun and
the subgiant model, respectively. We note that the maximum of ${\cal M}$ is about 0.7
and changes little from the ZAMS to the present sun. Using ${\cal M}$, we can define the 
turbulent kinetic energy per unit mass $\chi$ as
\begin{equation}
  \chi = \case{1}{2} {\cal M}^2 v_s^2.
\end{equation}
The turbulent contribution to the entropy is
\begin{equation}
  S_{\mbox{\scriptsize{turb}}} = \chi/T,
\end{equation}
where $T$ is the gas temperature.

\vspace{3mm}
\centerline{\epsfysize=6.5cm \epsfbox{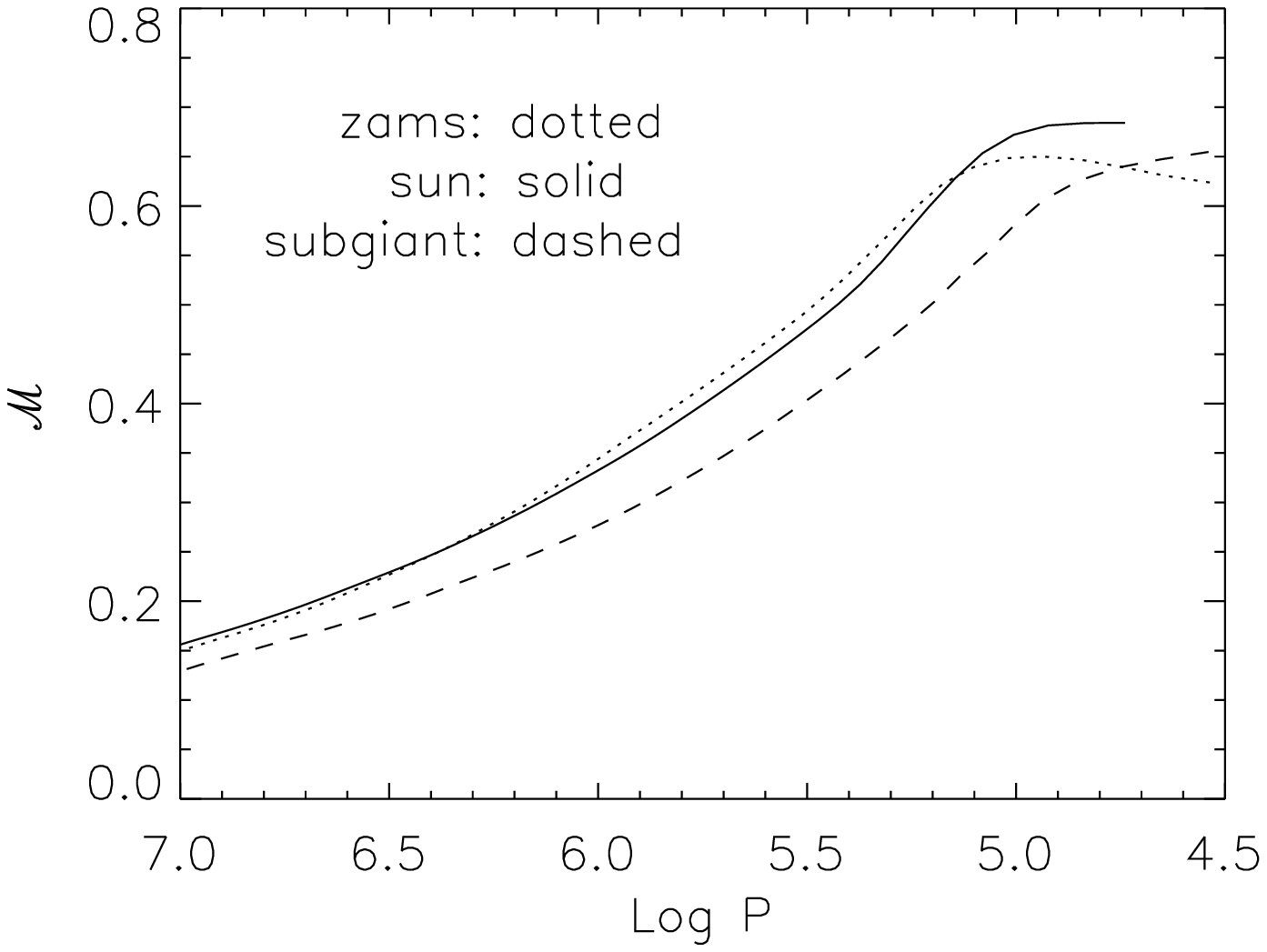}}
\figcaption[f1.eps]{
Turbulent Mach number as a function of depth at 3 evolutionary stages.
\label{mach}
}

\vspace{3mm}

\vspace{3mm}
\centerline{\epsfysize=6.5cm \epsfbox{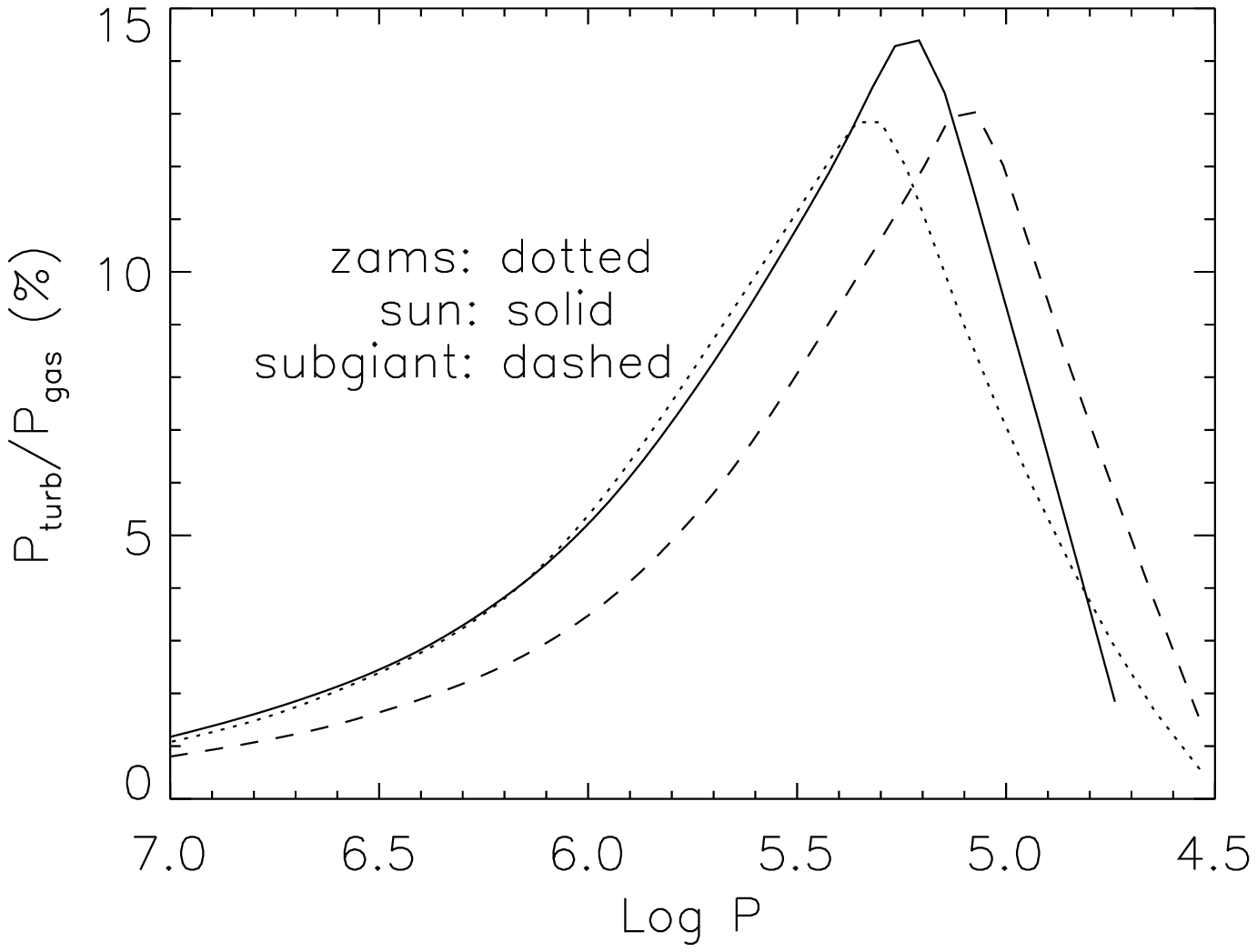}}
\figcaption[f2.eps]{
Ratio of turbulent to total pressure in the outer layers as a function of
depth at 3 evolutionary stages.
\label{pturb}
}

\vspace{3mm}

Turbulence in the stratified layers of the solar convection zone is not isotropic.
For convenience, we define the parameter $\gamma$ to reflect the anisotropy of turbulence,
\begin{equation}
  P_{\mbox{\scriptsize{turb}}}= (\gamma-1)\rho\chi, \label{eq:pturb}
\end{equation}
where $\rho\chi$ is the turbulent kinetic energy density. Since $P_{\s{turb}}=\rho {v''_z}^2$,
we can relate $\gamma$ to the turbulent velocity as follows:
\begin{equation}
  \gamma = 1 + 2(v''_z/v'')^2.
\end{equation}
$\gamma = 5/3$ when turbulence is isotropic ($v''_z=v''_x=v''_y$); $\gamma=3$ or
$\gamma=1$ when turbulence is completely anisotropic ($v''_z=v''$ or $v''_z=0$,
respectively). The physical meaning of $\gamma$ is the specific heat ratio due to 
turbulence. This affects the radial turbulent pressure distribution. Fig.~\ref{pturb} 
shows $P_{\s{turb}}$,  in which the radial turbulent pressure is scaled with  the gas pressure, 
$P_{\mbox{\scriptsize{gas}}}$. The total pressure is defined as
\begin{equation}
P_T=P_{\mbox{\scriptsize{gas}}}+P_{\mbox{\scriptsize{rad}}}+P_{\mbox{\scriptsize{turb}}}.
\end{equation}
Note that $P_{\mbox{\scriptsize{turb}}}/P_{\mbox{\scriptsize{gas}}} = (v''_z/v_s)^2$ since
$v_s=(P_{\mbox{\scriptsize{gas}}}/\rho)^{1/2}$.

\vspace{3mm}
\centerline{\epsfysize=6.5cm \epsfbox{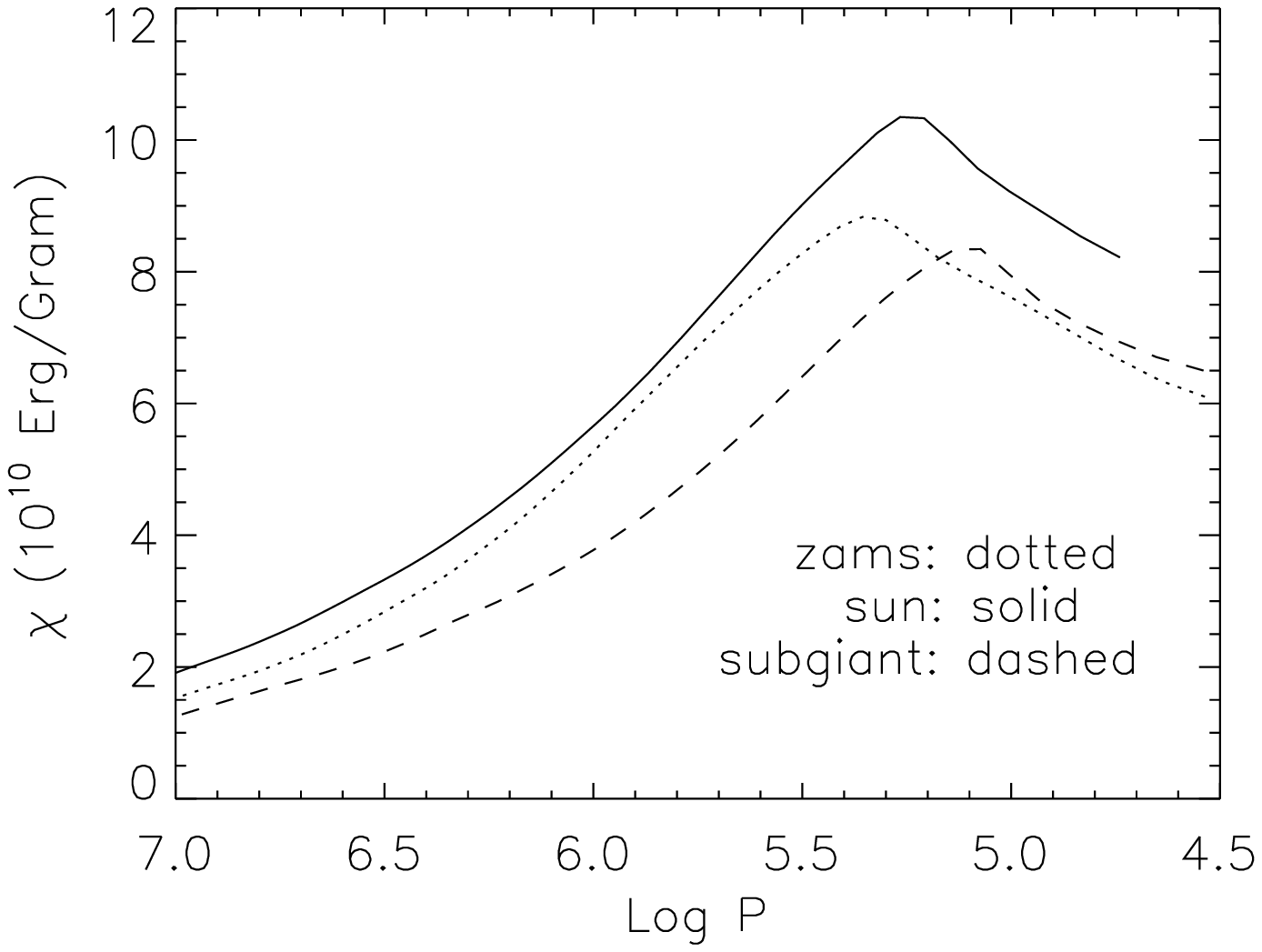}}
\figcaption[f3.eps]{
Turbulent kinetic energy per unit mass as a function of
depth at 3 evolutionary stages.
\label{chi}
}
\vspace{3mm}

\vspace{3mm}
\centerline{\epsfysize=6.5cm \epsfbox{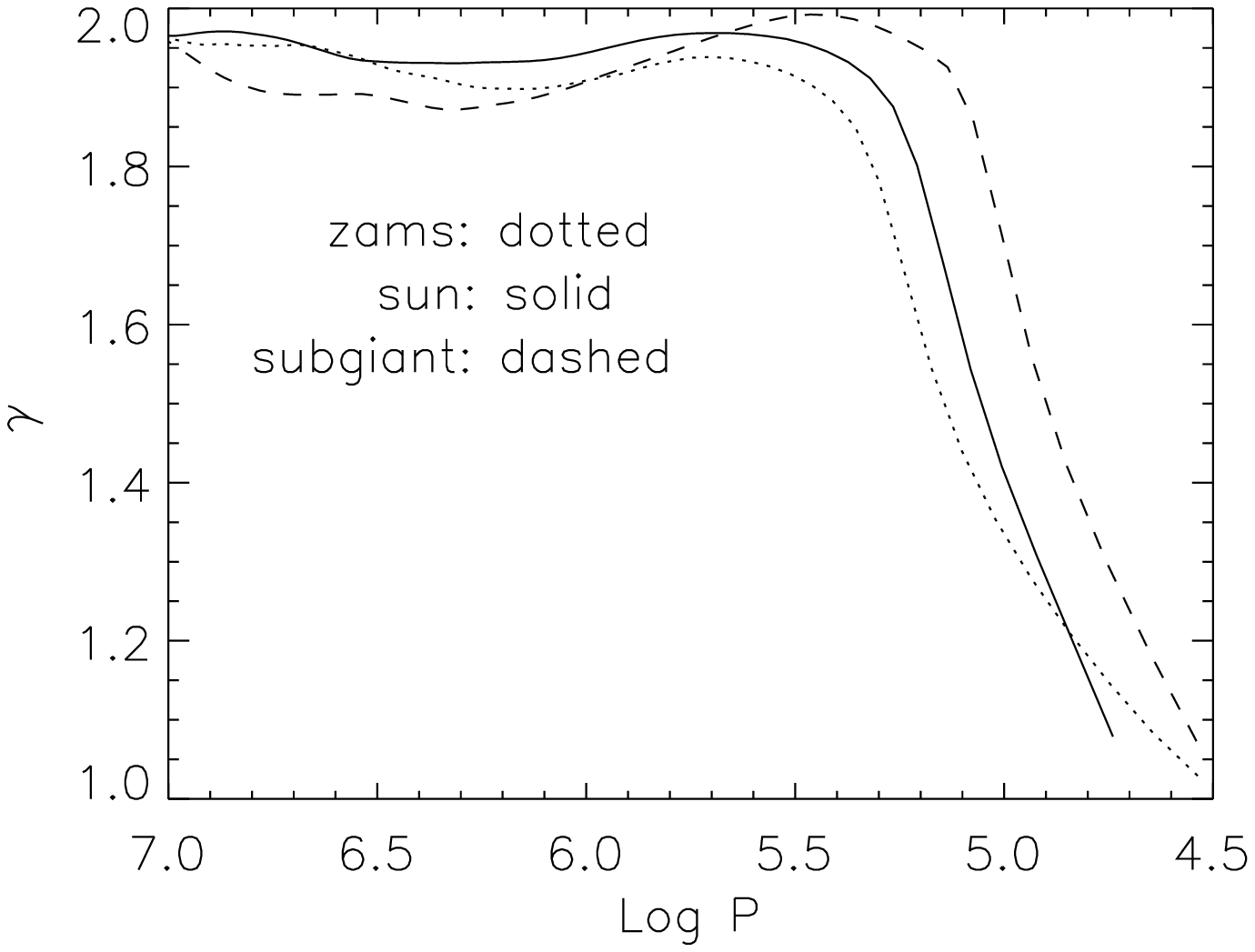}}
\figcaption[f4.eps]{
Specific heat ratio due to turbulence as a function of
depth at 3 evolutionary stages.
\label{gamma}
}
\vspace{3mm}

The turbulent contribution to the pressure, kinetic energy and entropy,
can all be expressed in terms of $\chi$ and $\gamma$. Therefore, turbulence can be 
parameterized by these two parameters. Figs.~\ref{chi} and \ref{gamma} show their  
variations as functions of depth.

\section{Convective temperature gradients with the turbulent 
velocities}\label{s3}

Abbett et al. (1997) tried to include turbulence in solar modeling within the
framework of MLT by using simulated pressure and density in calculating
the temperature gradient in the convection zone. As they pointed out, such an 
application of MLT is not self-consistent. Lydon and Sofia (1995) developed a 
self-consistent method to include magnetic fields in calculating the convective 
temperature gradients within the framework of MLT. Then, Lydon, Guenther and Sofia (1996) used
it to successfully explain the observed variation of solar p-modes with the
solar cycle. Recently, Li and Sofia (2001) have updated this method to reproduce
the observed cyclic variations of all solar global parameters such as solar
luminosity, solar effective temperature and solar radius. We briefly show how
the same method can be used to calculate the influence of turbulence on the temperature 
gradients in the convection zone.

\subsection{Turbulent variables}

Introducing $\chi$, and $\gamma$, computed from the 3D simulations,
the equation of state becomes
\begin{equation}
  \rho=\rho(P_T,T,\chi,\gamma).
\end{equation}
Including the turbulent 
kinetic energy, the first law of thermodynamics becomes
\begin{eqnarray}
  dQ_T &=& dU+PdV+d\chi \nonumber\\
       &=& dU_T + [P_T-(\gamma-1)(\chi/V)]dV, \label{eq:1stlaw}
\end{eqnarray}
where $U_T = U + \chi$, is the total internal energy per unit mass and  
$V=1/\rho$ is the volume per unit mass.

\subsection{Convective stability criterion}

The convective stability criterion can still be expressed by the difference
between density derivatives of a mass element and its surroundings:
\begin{equation}
  (d\rho/dr)_e - (d\rho/dr)_s > 0.
\end{equation}
However, since we are using $P_T$, $T$, $\chi$ and $\gamma$ as the independent
variables, we have
\begin{equation}
  d\rho/\rho = \mu d P_T/P_T - \mu' dT/T - \nu d\chi/\chi - \nu' d\gamma/\gamma,
\end{equation}
where
\[
\begin{array}{ll}
\mu = \left(\p{\ln\rho}{\ln P_T}\right)_{T,\chi,\gamma} &
\mu' = -\left(\p{\ln\rho}{\ln T}\right)_{P_T, \chi, \gamma}  \\
\nu = -\left(\p{\ln\rho}{\ln \chi}\right)_{P_T, T, \gamma}  &
\nu' = -\left(\p{\ln\rho}{\ln \gamma}\right)_{P_T, T, \chi}   \\
\end{array}
\]
As a result, the stability criterion becomes
\begin{equation}
  \nabla_{\mbox{\scriptsize{rad}}} < \nabla_{\mbox{\scriptsize{ad}}}-A_m.
\end{equation}
In this expression, $A_m$ reflects the direct influence of turbulence, defined by
\begin{equation}
A_m = (\nu\nabla_{\chi}+\nu'\nabla_{\gamma}) \nabla_{\mbox{\scriptsize{ad}}}/\mu,
\end{equation}
where we have assumed $(d\chi/dr)_e=(d\chi/dr)_s$ and
$(d\gamma/dr)_e=(d\gamma/dr)_s$, while
\[
\begin{array}{ll}
\nabla_{\chi} = \left(\p{\ln \chi}{\ln P_T}\right)_s &
\nabla_{\gamma} = \left(\p{\ln \chi}{\ln P_T}\right)_s \\
\nabla_{\mbox{\scriptsize{rad}}}=\frac{3}{16\pi a cG}\frac{\kappa L_r
P_T}{M_rT^4} &
\nabla_{\mbox{\scriptsize{ad}}} = \frac{P_T\delta}{\rho T c_p}
\end{array}
\]
The other symbols have the usual meanings.

\subsection{Convective temperature gradients}

\subsubsection{Flux conservation with turbulence}

The convective temperature gradient $\nabla_{\mbox{\scriptsize{conv}}}$ is
determined by the requirement that the total energy flux
$F_{\mbox{\scriptsize{total}}}$ equals the sum of the radiative flux
$F_{\mbox{\scriptsize{rad}}}$ and the convective flux 
$F_{\mbox{\scriptsize{conv}}}$,
\begin{equation}
   F_{\mbox{\scriptsize{total}}} = F_{\mbox{\scriptsize{rad}}} +
F_{\mbox{\scriptsize{conv}}}. \label{eq:flux}
\end{equation}
The total flux at any given layer in the star is determined by the photon
luminosity $L_r$,
\begin{equation}
  F_{\mbox{\scriptsize{total}}} = \frac{L_r}{4\pi r^2} 
     = \frac{4acG}{3}\frac{T^4M_r}{\kappa P_T  r^2}\nabla_{\mbox{\scriptsize{rad}}}.
   \label{eq:ftot}
\end{equation}
The radiative flux is determined by the convective temperature gradient:
\begin{equation}
  F_{\mbox{\scriptsize{rad}}} =  \frac{4acG}{3}\frac{T^4M_r}{\kappa P_T r^2}
     \nabla_{\mbox{\scriptsize{conv}}}. \label{eq:frad}
\end{equation}
The convective flux is determined by the convective velocity
$v_{\mbox{\scriptsize{conv}}}$ and the heat excess $DQ_T$:
\begin{equation}
  F_{\mbox{\scriptsize{conv}}} = \rho v_{\mbox{\scriptsize{conv}}}DQ_T. \label{eq:fconv}
\end{equation}
When the convective velocity is much smaller than the sound speed of the medium, 
the process can be considered to be of constant pressure. In this case, the heat
excess can be obtained from the first law given by Eq.~(\ref{eq:1stlaw}):
\begin{equation}
  DQ_T = c_pDT+\left[\frac{P_T\mu'\nu}{\rho\mu\chi}+1\right]D\chi 
   + \frac{P_T\mu'\nu'}{\rho\mu\gamma}D\gamma.
\end{equation}

\subsubsection{Mixing length approximation}

Using the mixing length approximation, $DT$, $D\chi$ and $D\gamma$ can be expressed
by the mixing length $l_m$ as follows:
\begin{eqnarray}
  DT/T &=& (1/T)\partial(DT)/\partial r (l_m/2) \nonumber\\
       &=& (\nabla_{\mbox{\scriptsize{conv}}}-\nabla_e)(l_m/2)(1/H_p),\nonumber \\
  D\chi/\chi &=& (1/\chi)\partial (D\chi)\partial r(l_m/2)
             = 0, \\
  D\gamma/\gamma &=& (1/\gamma)\partial (D\gamma)\partial r(l_m/2) = 0, \nonumber
\end{eqnarray}
where $H_p=-P_T(dr/dP_T)$ is the pressure scale height.
In order to determine $v_{\mbox{\scriptsize{conv}}}$, the MLT assumes that 
half of the work done by half of the radial buoyancy force acted over half the 
mixing length goes into the kinetic energy of the element 
($v_{\mbox{\scriptsize{conv}}}^2/2$). Since the radial buoyancy force per unit 
mass is related to the density difference by:
\begin{equation}
  k_r=-g(D\rho/\rho),
\end{equation}
and since the process is in pressure equilibrium, we obtain
\begin{equation}
  v_{\mbox{\scriptsize{conv}}}^2 
   = \mu'(\nabla_{\mbox{\scriptsize{conv}}}-\nabla_e)\frac{l_m^2g}{8H_p},
\end{equation}
where $g=GM_r/r^2$ is the gravitational acceleration.

An additional relation is required to close the MLT:
\begin{equation}
  (dQ_T/dr)_e = \mbox{(radiative losses) + \{change of $\chi$} \},
\end{equation}
which can be expressed as:
\begin{eqnarray}
  (2acT^3)/(\rho c_p v_{\mbox{\scriptsize{conv}}})(\omega/(1+\case{1}{3}\omega^2)
    (\nabla_{\mbox{\scriptsize{conv}}}-\nabla_e) \nonumber \\
  =(\nabla_e-\nabla_{\mbox{\scriptsize{ad}}}) +
(\nabla_{\mbox{\scriptsize{ad}}}/\mu)(\nu\nabla_{\chi}+\nu'\nabla_{\gamma}),
\label{eq:loss}
\end{eqnarray}
where $\omega=\kappa\rho l_m$.

\subsubsection{Result}

Solving Eqs.~(\ref{eq:flux}) and (\ref{eq:loss}), we obtain
\begin{equation}
  \nabla_{\mbox{\scriptsize{conv}}} = \nabla_{\mbox{\scriptsize{ad}}} +
(y/V\gamma_0^2C)(1+y/V) - A_m,
\end{equation}
where $y$ is the solution of the following equation
\begin{equation}
  2Ay^3+Vy^2 + V^2y-V =0. \label{eq:cubic}
\end{equation}
$A$, $\gamma_0$, $C$, and $V$ are defined by
\begin{eqnarray*}
   A &=& (9/8)[\omega^2/(3+\omega^2)], \\
   \gamma_0 &=& [(c_{\rm p}\rho)/(2acT^3)][(1+\case{1}{3}\omega^2)/\omega],  \\
   C &=& (g/l_m^2\mu')/8H_p, \\
   V&=& 1/[\gamma_0 C^{1/2}(\nabla_{\mbox{\scriptsize{rad}}}-\nabla_{\mbox{\scriptsize{ad}}}+A_m)^{1/2}].
\end{eqnarray*}
From these formulas it can be seen that the effect of turbulence on the temperature gradient 
in the convection zone can be taken into account by modifying the adiabatic gradient:
\begin{equation}
  \nabla'_{\mbox{\scriptsize{ad}}} = [1-(\nu\nabla_\chi+\nu'\nabla_\gamma)/\mu]
    \nabla_{\mbox{\scriptsize{ad}}}.
\end{equation}
Therefore, it is easy to include this effect in the standard stellar structure codes.

\section{Calibrated solar models}\label{s4}

\subsection{Standard solar model}

For the purpose of comparison, we construct a standard solar model with the
Yale Stellar Evolution Code. The OPAL opacities tables (Iglesias \&\ Rogers
1996) are used together with the low-temperature opacities from Kurucz (1991).
The equation of state is taken from Rogers, Swenson, \&\ Iglesias (1996). When
out of the table, the Yale standard implementation with the Debye-H\"{u}kel
correction is used (Guenther et al.1992). Helium and heavy element diffusion
processes are included in the model. Heavy element diffusion is implemented by
assuming that all heavy elements diffuse with the same velocity as fully ionized
iron (Guenther \& Demarque 1997). The model atmosphere is constructed using the 
empirical Krishna-Swamy (KS) $T-\tau$ relation \cite{GDKP92}. The solar model 
is evolved from the zero-age main sequence to the current solar age. The mixing-length ratio
$\alpha$ and the helium content $Y$ have been adjusted in the usual way so
as to match the solar luminosity, radius, and the ratio of heavy elements to
hydrogen ($Z/X=0.0230$) at the surface of the model (Grevesse \&\ Sauval 1998).
These are obtained by choosing ($Z_0,X_0$) and $\alpha$ to  be
($0.0188,0.7091724$) and $2.135772$. The standard (or reference) solar model is
abbreviated as the SSM.

\subsection{Solar model with turbulent pressure alone}

The simplest way to taken into account turbulence in solar modeling is to
include turbulent pressure (or Reynolds stress) alone, as done by many authors
(e.g., Balmforth 1992a). In this case, only the hydrostatic equilibrium equation
needs to be modified as follows (method 1):
\begin{equation}
    \p{P}{M_r} = - \frac{GM_r}{4\pi r^4}(1+\beta),
\end{equation}
where $P=P_{\mbox{\scriptsize{gas}}}+P_{\mbox{\scriptsize{rad}}}$, and
\begin{equation}
  \beta = \left(\frac{2P_{\mbox{\scriptsize{turb}}}}{\rho g
r}-\p{P_{\mbox{\scriptsize{turb}}}}{P}
    \right) \left(1+\p{P_{\mbox{\scriptsize{turb}}}}{P} \right)^{-1}.
\label{eq:hydrostatic}
\end{equation}
Here $2P_{\mbox{\scriptsize{turb}}}/(\rho g r)$ originates from the spherical
coordinate system adopted, representing a kind of geometric effect. The
equations that govern the envelope integrations also need to be changed accordingly.

We implement this case by modifying the Yale Stellar Evolution Code and obtain a
nonstandard model in the same way we obtain the standard model. We assume
$P_{\mbox{\scriptsize{turb}}}$, set equal to its value for the present sun, does 
not change from the ZAMS to the present age of the sun. The adjustable parameters 
now are fixed as ($Z_0,X_0,\alpha$)=($0.0188,0.7092889,2.138190$). This model is 
called the Pressure Solar Model (PSM).

\subsection{Solar model with $\chi$ and $\gamma$ as independent variables}

The form of the continuity equation and the equation of transport of energy by radiation
is not affected by turbulence. The hydrostatic equation includes a
Reynolds stress term due to turbulence
\begin{equation}
    \frac{\partial P}{\partial r} = - \frac{GM_r}{r^2}\rho -
\frac{1}{r^2}\od{}{r}(r^2\rho v_rv_r),
\end{equation}
where $P=P_{\mbox{\scriptsize{gas}}}+P_{\mbox{\scriptsize{rad}}}$. Since the
last term can be rewritten as
$\partial P_{\mbox{\scriptsize{turb}}}/\partial r+2(\gamma-1)\chi/r$, this
equation becomes
\begin{equation}
  \p{P_T}{M_r} = - \frac{GM_r}{4\pi r^4} - \frac{2(\gamma-1)\chi}{4\pi r^3}.
  \label{eq:hstate}
\end{equation}
The last term on the right hand side of Eq.~(\ref{eq:hstate}) also embodies
the same spheric geometric effect as $2P_{\mbox{\scriptsize{turb}}}/(\rho g r)$
in Eq.~(\ref{eq:hydrostatic}). The energy conservation equation is also affected by
turbulence because the first law of thermodynamics should include the turbulent
kinetic energy
\begin{equation}
  \p{L_r}{M_r} = \epsilon - T\od{S_T}{t}, \label{eq:energy}
\end{equation}
where
\begin{equation}
  TdS_T = c_{\rm p}dT - \frac{\mu'}{\rho} dP_T + \left(1+\frac{P_T\mu'
\nu}{\rho\mu\chi}\right)d\chi +
                       \frac{P_T\mu'\nu'}{\rho\mu\gamma}d\gamma.
\label{eq:firstlaw}
\end{equation}
The equation of energy transport by convection,
\begin{equation}
  \p{T}{M_r} = - \frac{T}{P_T}\frac{GM_r}{4\pi r^4}
\nabla_{\mbox{\scriptsize{conv}}},
\label{eq:convection}
\end{equation}
does not change in form, but the convection temperature gradient, obtained in
the previous section, is different from that without turbulence. The equations
that govern envelope integrations also need to be changed accordingly. This method
will be referred below as method 2.

We implement this case by modifying the Yale Stellar Evolution Code. Once
again, we obtain a nonstandard model in the same way as we obtained 
the standard model. If we assume that $\chi$ and 
$\gamma$ do not change with time (letting them equal their values at the present age
of the sun), the adjustable parameters now must be set as $(Z_0,X_0,\alpha$) =
($0.0188,0.7092715,2.271540$). We use the spline interpolation of $\chi$ and
$\gamma$ given in \S\ref{s2} for their pressure dependence in the model calculations. 
We call this model the Energy Solar Model 1 (ESM1).

In order to investigate the evolutionary effects of $\chi$ and $\gamma$, 
we linearly interpolate between the two simulations that are closest to 
the evolutionary state of the model. In this case,
($Z_0,X_0,\alpha$) = ($0.0188,0.7092945,2.271462$) in order to match the
observed $L_{\sun}$, $R_{\sun}$, and ($Z/X$). We call this model as Energy Solar
Model 2 (ESM2).

\section{Influence of turbulence on the solar structure}\label{s5}

We shall now investigate how different methods for including turbulence in solar
modeling, affect the solar model structure.

\subsection{Measured by structural variables}

Fig.~\ref{ptdcp} depicts turbulence-induced relative changes (with respect to
the SSM) of the solar structural variables for the PSM ({\it dotted line}), 
ESM1 ({\it dashed line}) and ESM2 ({\it solid line} overlapped on the dashed line),
as functions of the logarithm of the total pressure with base 10. The changes for 
the pressure, temperature, and luminosity are calculated at the same radius coordinate, 
while the radius change is calculated at the same interior mass coordinate $M_r$.
From this figure it can be seen that
\begin{enumerate}
\item In all the cases turbulence affects the distribution of the 
pressure, temperature, and the (radiative and convective, rather than the total) luminosity 
around the SAL peak (specified by the dashed vertical line in the figure);
\item Both methods produce almost the same maximum pressure change of 15\%, but method 1 
produces a larger change for the temperature (8\% vs 4\%), radiative and convective luminosity 
(130\% vs 70\% and 100\% vs 80\%);
\item It is surprising that turbulence near the surface affects (slightly) the mass 
distribution (denoted by the radius change) in the core of the solar models.
\end{enumerate}

The increase of the radiative luminosity does not necessarily imply the 
increase of the convective gradient since the radiative flux depends on not only the 
convective gradient, but also the temperature $T$, the radiative opacity $\kappa$, and the 
pressure $P_T$, as expressed by Eq.~\ref{eq:frad}. Similarly, the decrease of the convective 
luminosity does not necessarily imply the increase of the convective gradient since the 
convective flux depends on the specific heat at constant pressure $c_p$, density $\rho$, 
temperature $T$, convective velocity $v_{\s{conv}}$, and the mixing length $l_m$:
\begin{equation}
  F_{\s{conv}}=\frac{4 c_p \rho T}{g\mu'l_m}v_{\s{conv}}^3.
\end{equation}
Fig.~\ref{dkvhc} shows how $\rho$, $\kappa$, $v_{\s{conv}}$, $l_m$, and $c_p$ in the PSM and ESM 
change with respect to the SSM. The increase of the radiative luminosity below the SAL peak is 
mostly generated by the decrease of the radiative opacity caused by the decrease of the 
temperature and density. The decrease of the convective luminosity above the SAL peak is mostly 
generated by thedecrease of the convective velocity.

Figs.~\ref{ptdcp} and \ref{dkvhc} show that the temperature, density and total pressure decrease 
when the turbulent pressure is included. The decrease of the total pressure is possible since 
what supports gravity in the solar interior is the total pressure gradient, not the pressure 
itself. Besides, the gas pressure may decrease in order to maintain hydrodynamic equilibrium when 
turbulence provides a turbulent pressure. Consequently, the gas density and/or the gas
temperature should decrease. However, the two methods make difference here: the decrease of the 
temperature (density) in the ESM is smaller (larger) than that in the PSM. The cause is that the 
mixing length in the ESM increases near the SAL peak, but the mixing length in the PSM decreases 
near the SAL peak. This implies that the transport of energy by convection near the peak in the ESM 
is more efficient than that in the PSM. As a result, different SAL structures are expected, as addressed 
in the next subsection.

\vspace{3mm}
\centerline{\epsfysize=10.5cm \epsfbox{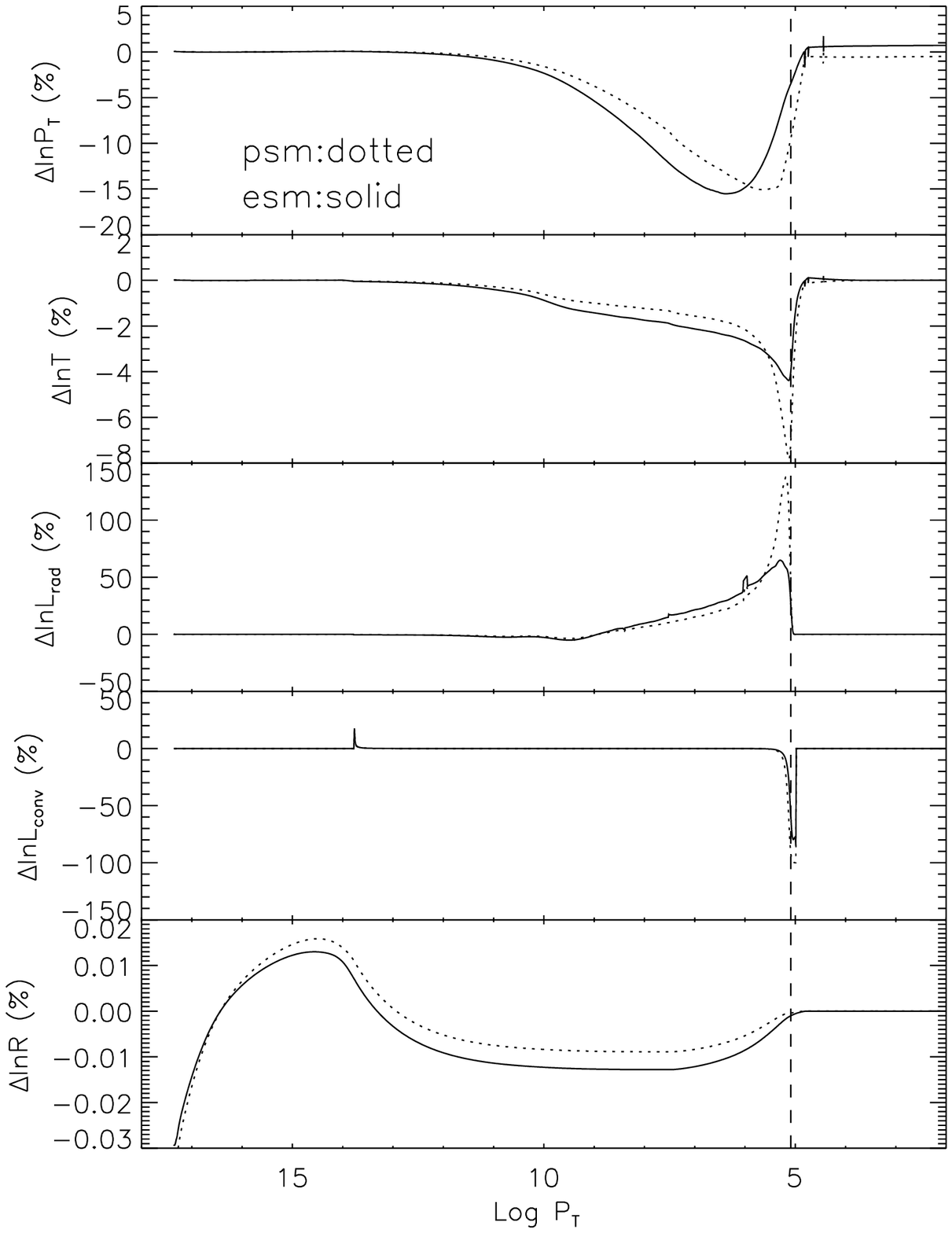}}
\figcaption[f5.eps]{
Turbulence-induced relative changes of the solar structural variables (pressure, temperature, radiative luminosity, convective luminosity, and radius) for the PSM and ESM with respect to the SSM. The vertical line 
indicates the location of the SAL peak of the SSM: $\mbox{Log} P_{\s{T}}=5.09$.
\label{ptdcp}
}
\vspace{3mm}

The comprehensive effect of turbulence on solar structure manifests itself in
the adiabatic sound speed
\begin{equation}
  \Delta\ln C = \case{1}{2} (\Delta\ln \Gamma_1 + \Delta\ln P_T - \Delta\ln\rho),
\end{equation}
where $\Gamma_1$ is the first adiabatic exponent defined by
\begin{equation}
  \Gamma_1 = \left(\p{\ln P_T}{\ln \rho}\right)_{S,\chi,\gamma}
\end{equation}
when turbulence is modeled by $\chi$ and $\gamma$. Fig.~\ref{biggamma} shows
these four quantities for the ESM and PSM. Obviously, the pressure and density 
changes contribute to the change of the adiabatic sound speed, as does the 
change of the first adiabatic exponent. This shows that the thermodynamic 
properties of the solar matter changes when turbulence is present, as expected.
Nevertheless, the two methods produce a difference for the first adiabatic exponent 
around the SAL peak.

\vspace{3mm}
\centerline{\epsfysize=10.5cm \epsfbox{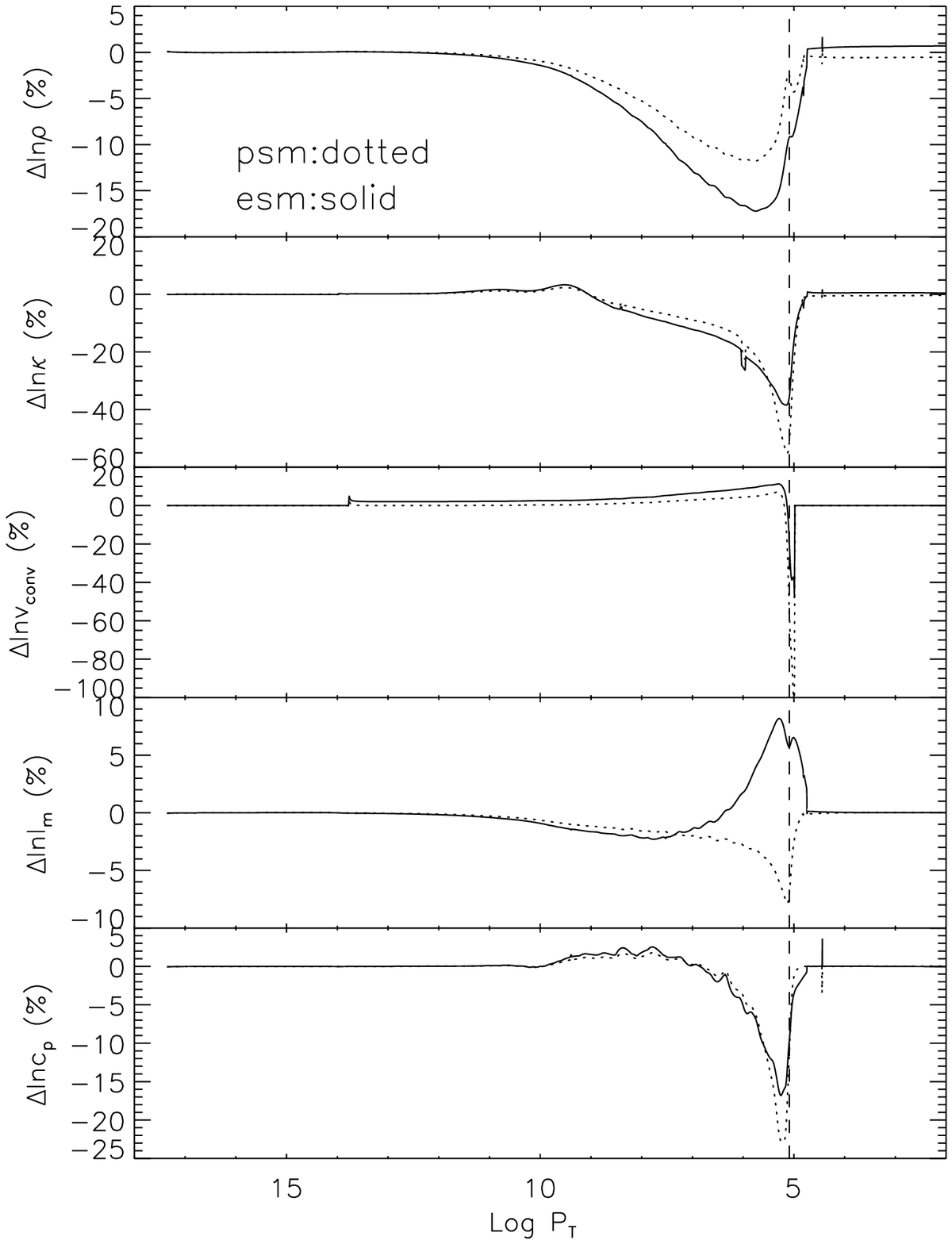}}
\figcaption[f6.eps]{
Realtive change of the density, opacity, convective velocity, mixing length, and specific heat at the constant pressure in the PSM ({\it dotted}), and ESM ({\it solid}) with respect to the SSM.
\label{dkvhc}
}
\vspace{3mm}

\vspace{3mm}
\centerline{\epsfysize=9cm \epsfbox{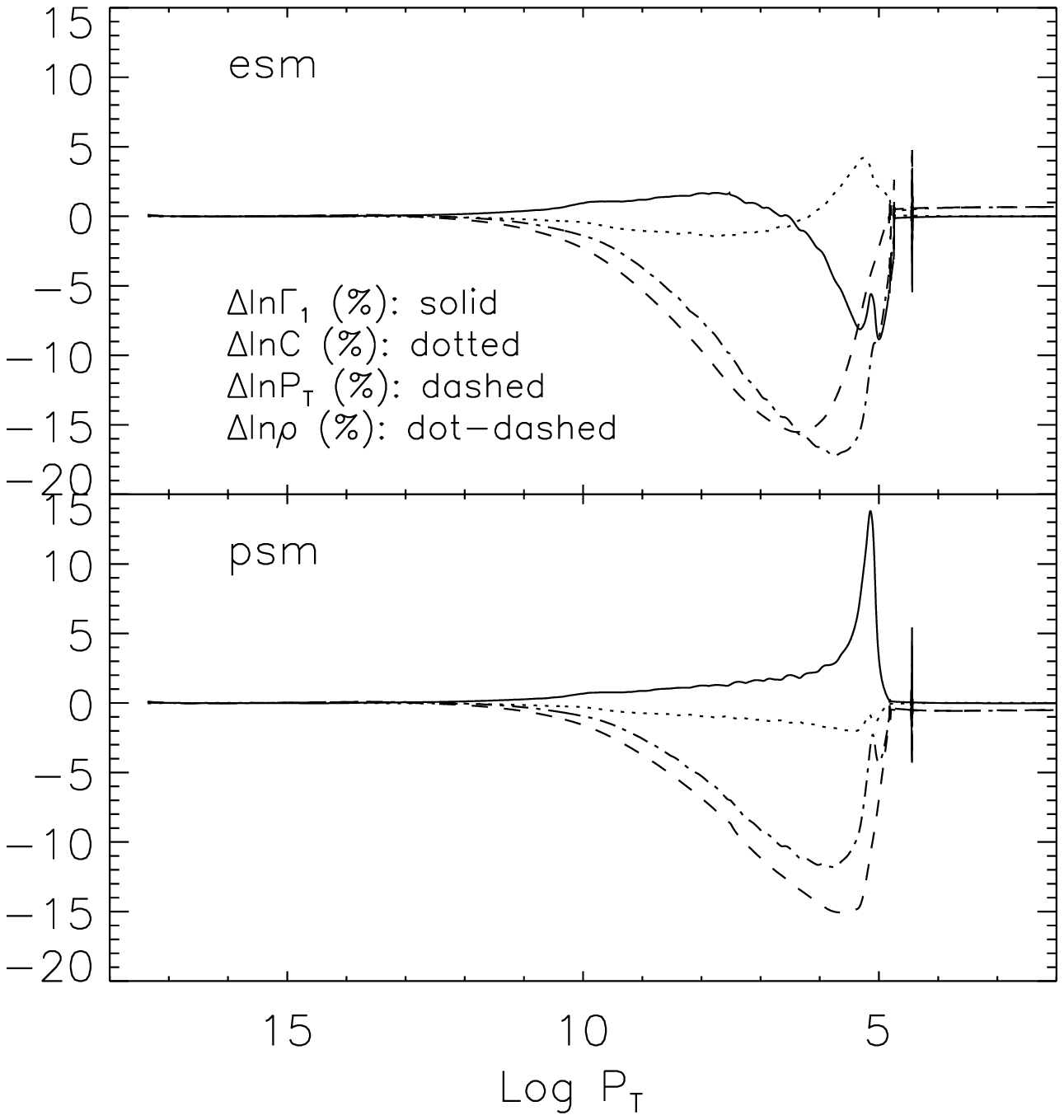}}
\figcaption[f7.eps]{
Relative change of the adiabatic sound speed and its contributors in the ESM and PSM.
\label{biggamma}
}
\vspace{3mm}

Another feature that we can see from Figs.~\ref{ptdcp}-\ref{biggamma} [the relative change 
for any variable $X$ is defined as follows: $(X_{\s{model}}-X_{\s{ssm}})/X_{\s{ssm}}$] is that
although turbulence is restricted to the highly superadiabatic layer ($\log P_T\in
[4.6,7]$, see Figs.~\ref{mach}-\ref{gamma}), its influence extends deeply into
the solar interior for the ESMs. For example, we still see some influence near
the base of the convective zone at $\log P_T\sim 13$. This is a natural
consequence of continuity.

\subsection{Measured by superadiabaticity}

The superadiabaticity ($\nabla - \nabla_{\mbox{\scriptsize{ad}}}$) as a function of
logarithm of total pressure with base 10 is depicted in Fig.~\ref{ss}
for the SSM ({\it solid line}), and ESM1 ({\it dotted line}). The PSM has the same SAL as the SSM,
and ESM2 has the same SAL as the ESM1. The SAL peak of the SSM equals to 0.45, while that of 
the ESM equals to 0.40, about 11\% lower than that of the SSM.

The corresponding 3D simulations produce an SAL very similar to the 1D model.
The maximum of $\nabla-\nabla_{\rm ad}$ is about 0.4. The 3D solar surface simulations 
by Stein and Nordlund \cite{RCNST99} and Canuto's 1D turbulence models found a much 
higher peak of about 0.8. The simulations by Stein and Nordlund employ a hyperviscosity model 
for the subgrid scales, while our 3D simulations use the Smagorinsky 
model \cite{S63} with the smallest viscosity that was numerically stable 
($c_\mu = 0.2$, see below). If we increased the viscosity by an order of magnitude then the
SAL peak was similar to Stein and Nordlund's.

\vspace{3mm}
\centerline{\epsfysize=6.5cm \epsfbox{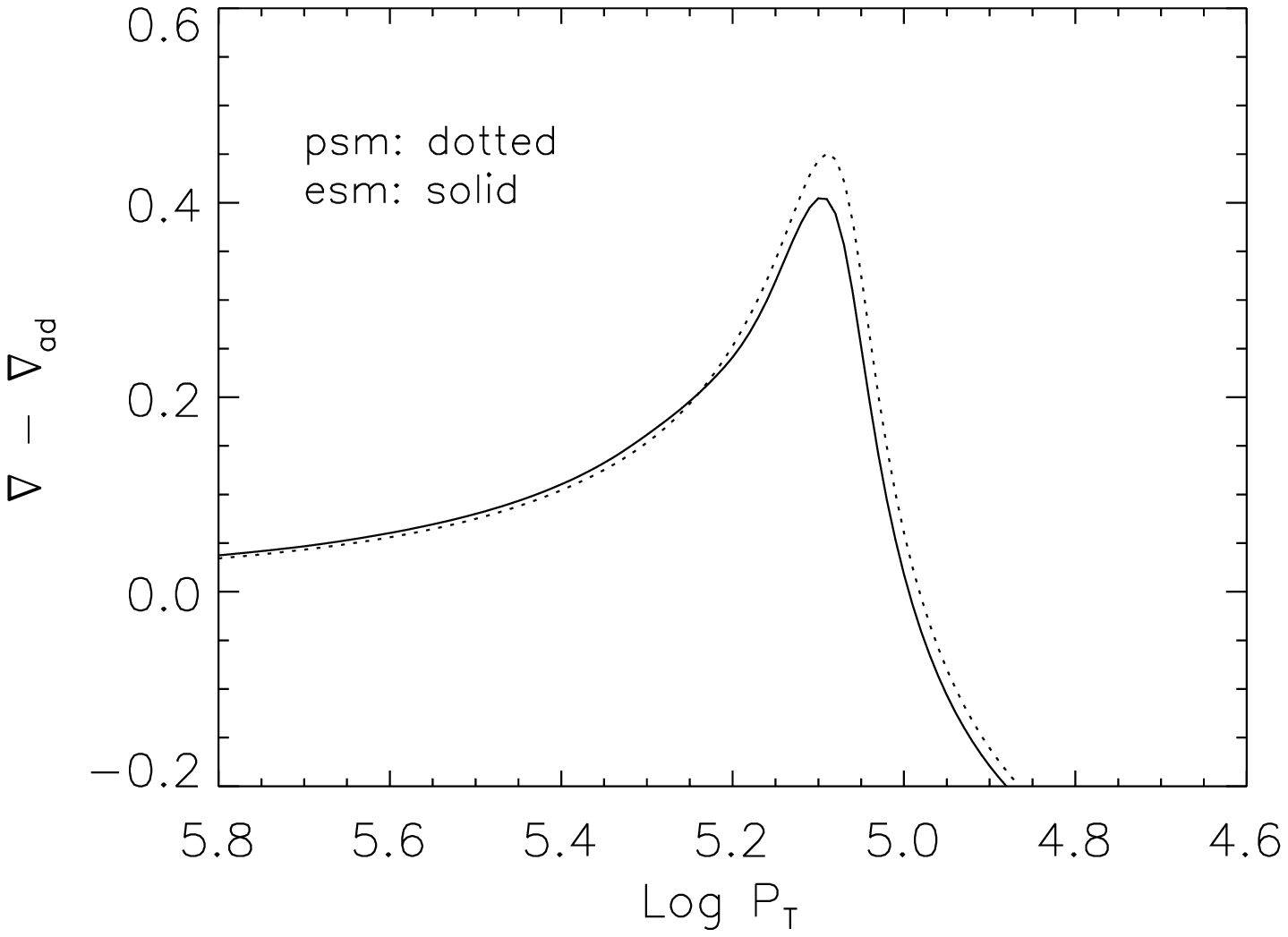}}
\figcaption[f8.eps]{
Structure of the highly superadiabatic layer (SAL). The SAL of the SSM is overlapped on that of the PSM.
\label{ss}
}
\vspace{3mm}

In order to understand these results, we note that the actual temperature gradient
is determined by the relative efficiency between the radiative and convective 
transport of heat since the total energy flux is fixed by the total luminosity, see 
Eq.~(\ref{eq:ftot}). Fig.~\ref{conv} shows the actual and adiabatic temperature gradients 
for the ESM and PSM, from which we can see that the SAL peak decrease in the ESM with respect 
to the SSM is mostly caused by the decrease of the actual gradient due to the inclusion of 
the turbulent kinetic energy. The maximum relative change of the convective (and adiabatic) gradient 
in the ESM is less than 10\%. In the MLT approach, the convective flux is proportional 
to the convective velocity, as shown by Eq.~(\ref{eq:fconv}). In 3D simulations, the total 
flux is equal to the radiative flux $F_{\s{rad}}$, plus the enthalpy flux $F_e$, plus the 
turbulent kinetic energy flux $F_k$. Near the SAL peak, $F_k\approx 0$. The enthalpy 
flux \cite{CS89} is proportional to the root mean square fluctuation of vertical 
velocity $v''_z$ defined in Eq.~(\ref{eq:rms}),
\begin{equation}
  F_e=\rho c_p \overline{v_z'T'} =\rho c_p C[v'_zT']v''_zT'',
\end{equation}
where $C[v'_zT']$ is the correlation coefficient between the temperature and vertical velocity 
fluctuations, and $T''$ is the root mean square fluctuation of temperature. Since $F_k\approx0$ 
near the SAL peak, the decrease of the convective gradient in the ESM and the 3D simulations 
implies an increase of the convective velocity $v_{\s{conv}}$ and the rms fluctuation of the 
vertical velocity just near the peak, as confirmed by Fig.~\ref{vv}. This shows that the SAL peak 
is sensitive to the turbulent velocity.

\vspace{3mm}
\centerline{\epsfysize=6.5cm \epsfbox{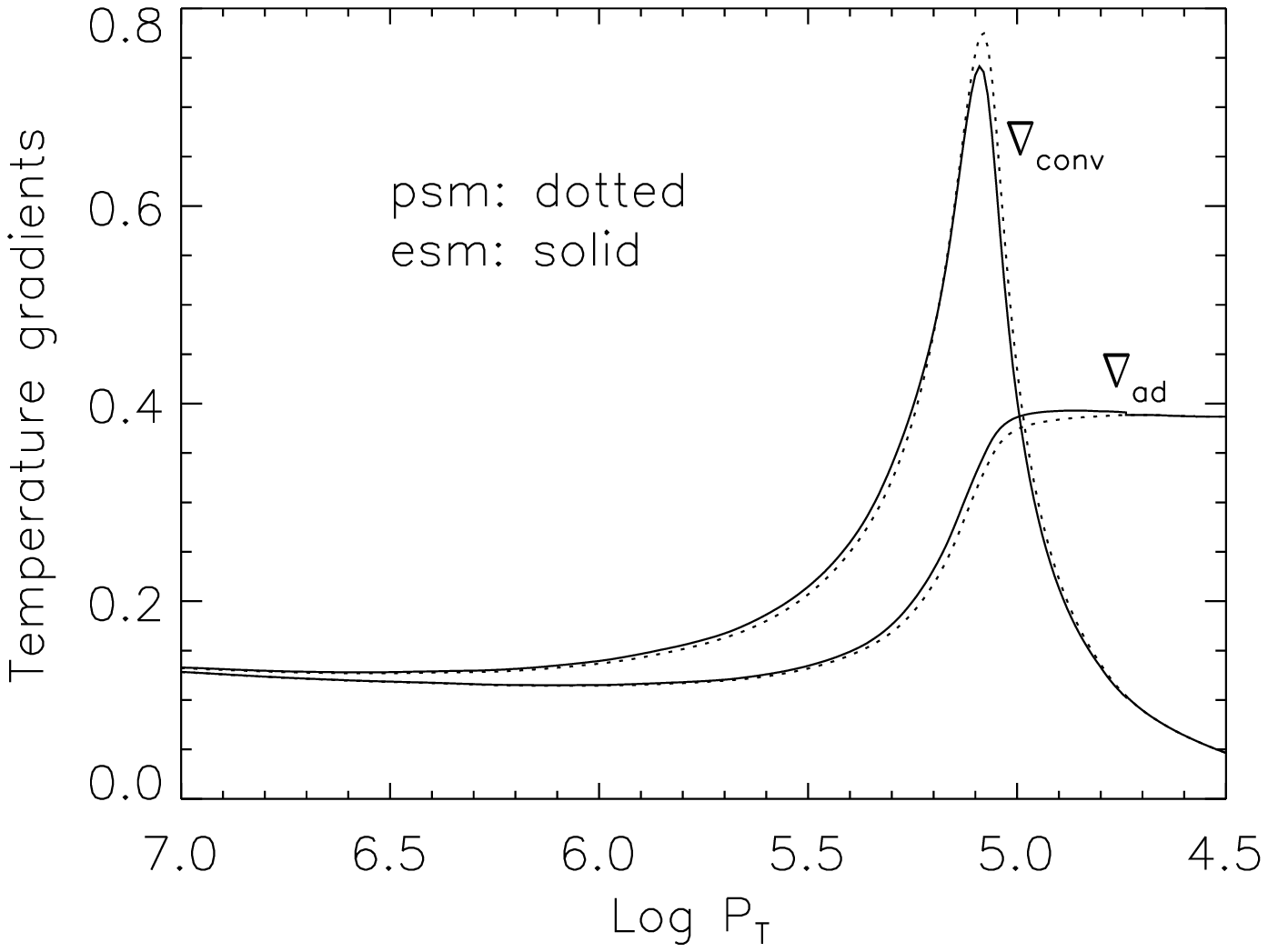}}
\figcaption[f9.eps]{
Convective and adiabatic temperature gradients. Those of the SSM and PSM are overlapped on each other.
\label{conv}
}
\vspace{3mm}

The most reliable method to determine the turbulent velocity is by direct 
numerical simulations (DNS) using the real solar kinematic viscosity $\nu$ \cite{LL87}. Since the 
number of degrees of freedom needed to represent 3D turbulent convection is proportional 
to Re$^{9/4}$, to resolve numerically all the scales in the solar convection zone (where 
the Reynolds number $\mbox{Re}\approx 10^{12}$) would require $10^{27}$ grid points \cite{C00}. 
However, the maximum number of grid points allowed by the present technology in DNS is o($10^9$). 
This forces us to use large eddy simulations (LES) by increasing the kinematic viscosity $\nu$ 
so that it represents the effects of Reynolds stresses on the unresolved or sub-grid 
scales (SGS). In the LES simulations, we used the SGS formula due to Smagorinsky (1963),
\begin{equation}
  \nu=(c_\mu\Delta)^2(2\sigma:\sigma)^{1/2}. \label{eq:sgs}
\end{equation}
The colon inside the brackets denots tensor contraction of the rate of strain $\sigma_{ij}
=(\nabla_i \bar{v_j}+\nabla_j \bar{v_i})/2$, $\Delta$ is the grid spacing, and $c_\mu$ is an 
adjustable dimensionless parameter. In the Smagorinsky model, $\nu\approx (c_\mu\Delta)^2u/L$, 
where the characteristic $u$ and length $L$ are estimated from the resolved motions. Consequently,
the Reynolds number reads
\begin{equation}
  \mbox{Re}=uL/\nu=(L/c_\mu\Delta)^2.
\end{equation}
In nondimensional units, $L\approx 1$, $c_\mu=0.2$ and $\Delta\approx 0.01$. 
So $\mbox{Re}\sim 10^6$. This number is still much smaller than the 
Reynolds number $10^{12}$ in the solar convection zone. Therefore, the viscosity must be 
overestimated, although this Reynolds number should be high enough for the fluid to be turbulent.
The fact that the standard MLT generates a lower convective velocity may imply that the MLT 
assumes a larger viscosity than the 3D simulations. This consideration disfavors those solar 
models with a higher SAL peak than that of the SSM.

\vspace{3mm}
\centerline{\epsfysize=6.5cm \epsfbox{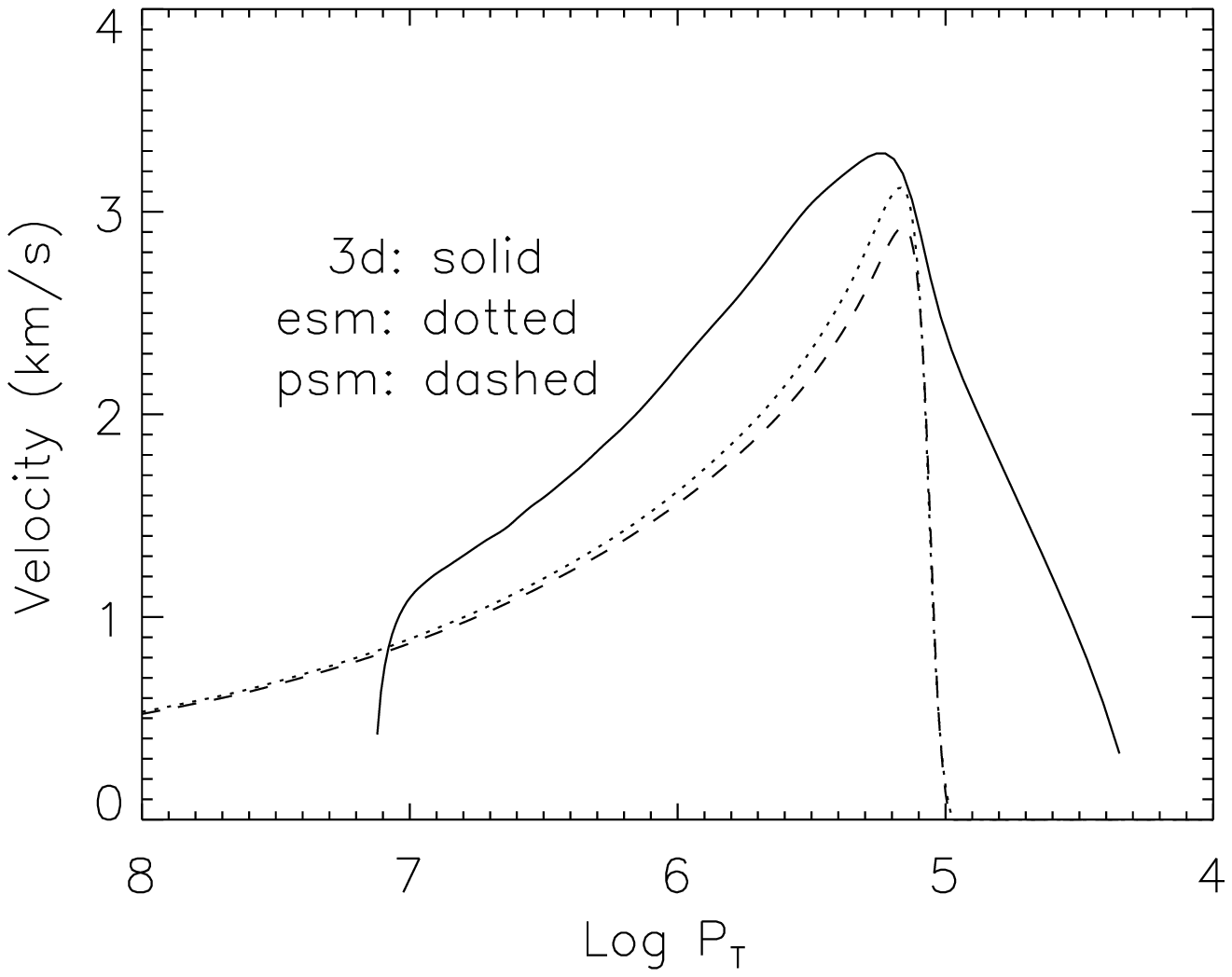}}
\figcaption[f10.eps]{Radial turbulent velocity and convective velocity as a function of depth (measured by
logarithm of pressure with base 10). Those of the SSM and PSM are overlapped on each other.
\label{vv}
}
\vspace{3mm}

It should be pointed out that we compare different solar models at the same radius coordinate while
we compare the SAL at the same pressure coordinate. Therefore, we observe a large difference for 
the convective velocity in Figs.~\ref{dkvhc} but a small (or no) difference between the ESM (PSM) and 
the SSM in Fig.~\ref{conv}.

\subsection{Measured by p-mode oscillation frequencies}

Our principal goal is to investigate how the treatment of turbulence in solar
modeling affects the computed model structure. Therefore, we do not include the
contribution of turbulence to the pulsation equations when we calculate p-mode
oscillation frequency differences caused by turbulence. We use Guenther's pulsation code
(1994) to calculate the p-mode frequencies under the adiabatic approximation,
for our SSM, PSM, ESM1 and ESM2, respectively.

Fig.~\ref{turbdif} shows the frequency differences (turbulent solar model,
including the PSM, ESM1 and ESM2, minus standard solar model) scaled by the mode
mass $Q_{\s{nl}}$ (e.g., Christensen-Dalsgaard et al. 1991). The frequency
differences between the PSM and SSM are comparable with Balmforth's
(Table 1, 1992a). Both models use turbulent pressure alone. The
frequency differencies between the EMS and SSM are comparable with
Rosenthal et al.'s (Figure 5, 1999).

In order to examine if the frequency differences shown in Fig.~\ref{turbdif} are
caused by the calibration of the solar models, we calculated the corresponding
uncalibrated models for the calibrated PSM and ESM2, respectively. Unlike the
calibrated model, the starting model for the uncalibrated model is the standard
solar model at the present age of the sun, instead of the ZAMS model. We switch on
the turbulence and evolve the model 10 timesteps with 1 year as the timestep length.
The frequency differences between calibrated and uncalibrated turbulent models
are less than 1 $\mu$Hz.

\vspace{3mm}
\centerline{\epsfysize=6.5cm \epsfbox{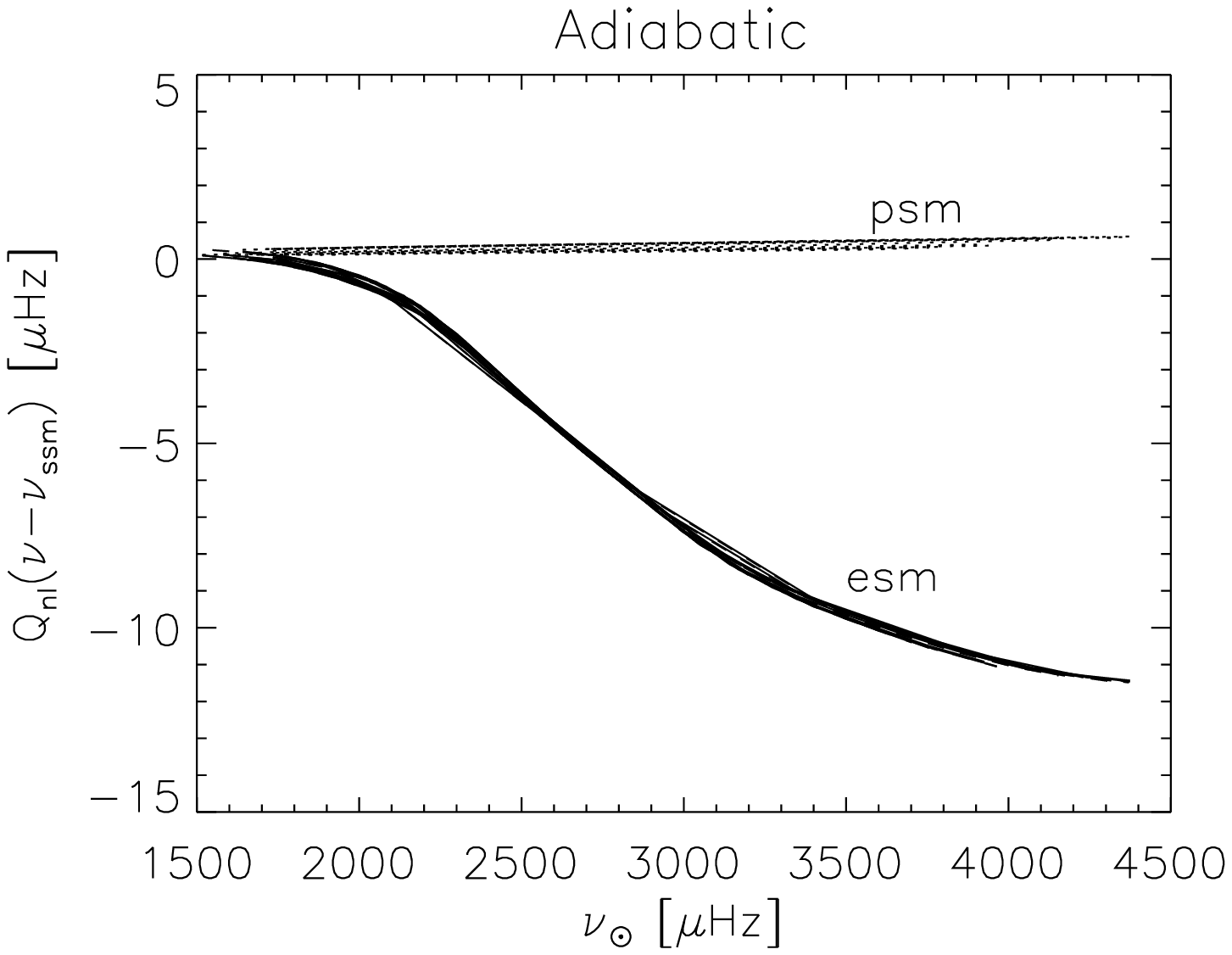}}
\figcaption[f11.eps]{
P-mode frequency difference diagrams. Turbulent model minus standard model, for
the turbulent pressure solar model (psm), and the solar model with the turbulent pressure 
and turbulent kinetic energy (esm1 and esm2). Plotted are the l = 0, 1, 2, 3, 4,
10, 20, ..., 100 p-modes.
\label{turbdif}
}
\vspace{3mm}

By calculating the frequency differences between the ESM1 and ESM2 models we can
find out the effect of temporal change of turbulence in the evolutionary
timescale of the sun. From Fig.~\ref{turbdif} we can see that this effect is
very small (less than 0.5 $\mu$Hz).

\vspace{3mm}
\centerline{\epsfysize=6.5cm \epsfbox{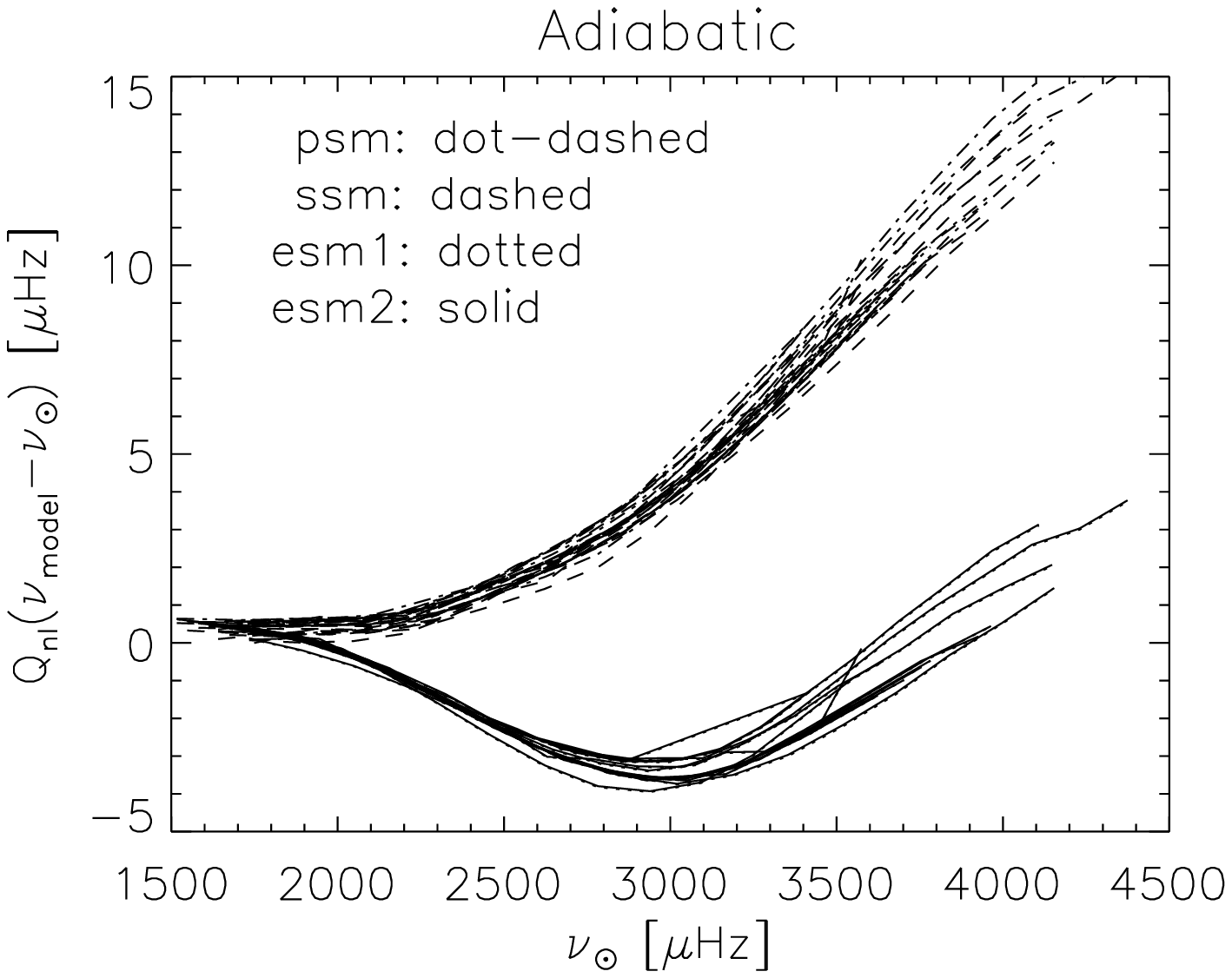}}
\figcaption[f12.eps]{
P-mode frequency difference diagrams, observation minus model, scaled by the
mode mass $Q_{nl}$, for the standard solar model (ssm), the turbulent pressure
solar model (psm), the solar model with fixed turbulent pressure and kinetic
energy (esm1), and the solar model with evolutionary turbulent pressure and
kinetic energy (esm2, almost overlap with esm1). Plotted are the l = 0, 
1, 2, 3, 4, 10, 20,...,100 p-modes.
\label{pmode}
}
\vspace{3mm}

As pointed out ealier, the PSM model is obtained by including turbulent pressure
alone in the solar modeling, while the ESMs are obtained by introducing the turbulent
variables $\chi$ and $\gamma$ to include both turbulent pressure and kinetic
energy. Therefore, the frequency differences between these two
kinds of models reflects the different treatments of 
turbulence in the solar modeling. Physically, the differences are caused by
the addition of turbulent kinetic energy. Fig.~\ref{turbdif} shows that the frequency
differences caused by turbulent kinetic energy are much larger than those caused
by turbulent pressure alone in size. Fig.~\ref{pmode} shows that the frequency
changes caused by turbulent kinetic energy make the computed model frequencies
match the solar data better than the SSM model, which is in agreement with
Rosenthal et al.'s result (see their Figures~1 and 6, 1999).

\section{Concluding remarks}\label{s6}

We have shown how different treatments of turbulence in solar modeling affect
the model structure within the framework of the standard mixing length theory.
The turbulent velocity is obtained from 3D numerical simulations of turbulence
in the highly  superadiabatic layer of the sun. When we introduce the turbulent 
kinetic energy per unit mass $\chi$ and the effective ratio of specific heats 
due to the turbulent perturbation $\gamma$, as independent thermodynamic variables, 
the resultant solar model is in agreement with the patched solar model, in which 
the simulated SAL replaces the original SAL. The frequency shift tends to match 
the observations better than the standard solar model. In contrast, when we use 
only the turbulent pressure, the turbulent effects are 
substantially underestimated (in the sense that the resultant p-mode frequency 
shift is much smaller).

Another difference between method 1 and 2 is that the SAL peak in the ESM is lower
than that of the SSM, but that of the PSM is the same as that of the SSM. The reason is that
the increase of the mixing length, in the vicinity of the SAL peak, by the turbulent 
kinetic energy, is more than double the reduction of the mixing length by the turbulent 
pressure (see panel 4 of Fig.~\ref{dkvhc}). The SAL produced by method 2 is consistent with
that of our 3D simulations, while method 1 does not change the SAL structure of the SSM.

Previously, the turbulent pressure was considered to play an important role in solar
modeling, but we have shown that this is not true: it is the turbulent kinetic
energy that is important. In fact, if we calibrate the solar model,
the elevation caused by the turbulent pressure $\Delta r$ vashishes.
Consequently,
\begin{equation}
  \frac{\delta\nu}{\nu} \approx \frac{\Delta
r/c_{\mbox{\scriptsize{ph}}}}{\int_0^{R_{\sun}} dr/C}
\end{equation}
vanishes, where $c_{\mbox{\scriptsize{ph}}}$ is the adiabatic sound speed at the
solar surface (photosphere). However, we obtain almost the same frequency
changes with and without calibrating the model radius to the solar radius at the
present age of the sun. Moreover, the solar model with the turbulent pressure
alone does not reproduce the p-model frequency shift obtained by Rosenthal et
al.'s patched model (1999).

To match the observed solar p-mode frequencies is not the only motivation to 
improve the solar model by including turbulence. More interesting is to generate 
a solar model that can excite the observed p-modes and damp the unobserved modes
simultaneously. Such an effort is in progress \cite{LRSDG01}. Overshooting 
may also play an important role in this regard.

\acknowledgements
This research was supported in part by a grant from the National Aeronautics and
Space Administration, and in part by Natural Science Foundation of China
(project 19675064, L. H. Li). Support from NASA grant
NAG5-8406 to Yale University, and from the National Science and
Engineering Research Council of Canada (D. B. Guenther) are also gratefully
acknowledged. Helpful comments from S. Basu are gratefully acknowledged.

{}

\end{document}